\documentclass[letterpaper, 10 pt, conference]{ieeeconf}  

\IEEEoverridecommandlockouts                              

\usepackage{graphicx}
\usepackage{epsfig}
\usepackage{color}
\usepackage{enumerate}
\usepackage[sort,compress]{cite}
\usepackage{graphicx}
\usepackage{subcaption}
\usepackage{romannum}
\usepackage{epstopdf}
\usepackage{balance}
\usepackage{bigints}
\usepackage{amssymb}
\usepackage{multirow}
\usepackage{amsmath}
\usepackage{hyperref}
\usepackage{graphicx}
\usepackage{epsfig}
\usepackage{mathptmx}
\usepackage{amsmath}
\usepackage{amssymb}
\usepackage[section]{placeins}
\usepackage{placeins}
\usepackage{amsmath}
\usepackage{multirow}
\usepackage{graphicx}
\usepackage[normalem]{ulem}
\usepackage{hyperref}
\usepackage{subcaption}
\usepackage{algorithm,tabularx}
\usepackage{algpseudocode}
\usepackage{amsfonts}
\usepackage{cases}
\usepackage{romannum}
\usepackage{epstopdf}
\usepackage{textcomp}
\usepackage{soul}
\usepackage{mathtools}
\usepackage{subcaption}
\usepackage{textcomp}
\usepackage{color}
\usepackage{url}%
\usepackage{makecell}
\usepackage{graphicx}
\usepackage[export]{adjustbox}
\usepackage{diagbox,hhline}
\usepackage[table]{xcolor}
\usepackage{balance}

\newcommand{\bea}{\begin{eqnarray}}
\newcommand{\eea}{\end{eqnarray}}
\newcommand{\beas}{\begin{eqnarray*}}
\newcommand{\eeas}{\end{eqnarray*}}
\newcommand{\leftm}{\left[\begin{array}}
\newcommand{\rightm}{\end{array}\right]}
\newcommand{\reals}{\mbox{$\mathbb R$}}

\newcommand{\ones}{\textbf{1}}

\newcommand{\mV}{\mathcal{V}}
\newcommand{\mN}{\mathcal{N}}

\DeclareMathOperator*{\argmax}{arg\,max}
\DeclareMathOperator*{\argmin}{arg\,min}
\DeclareMathOperator*{\argsup}{arg\,sup}
\DeclareMathOperator*{\softmax}{softmax}

\title{\LARGE \bf
Scaling Up Multiagent Reinforcement Learning for Robotic Systems:\\ Learn an Adaptive Sparse Communication Graph
}

\author{Chuangchuang Sun$^{}$, Macheng Shen,$^{}$ and Jonathan P. How$^{}$
\thanks{}
\thanks{$^{*}$Laboratory for Information and Decision Systems, Massachusetts Institute of Technology, 77 Massachusetts Ave, Cambridge, MA 02139. \newline
        {Emails: \tt \{ccsun1, macshen, jhow\}@mit.edu.}}%
}

\begin{document}

\maketitle
\thispagestyle{empty}
\pagestyle{empty}

\begin{abstract}
The complexity of multiagent reinforcement learning (MARL) in multiagent systems increases exponentially with respect to the agent number. This scalability issue prevents MARL from being applied in large-scale multiagent systems. However, one critical feature in MARL that is often neglected is that the interactions between agents are quite sparse. Without exploiting this sparsity structure, existing works aggregate information from all of the agents and thus have a high sample complexity. To address this issue, we propose an adaptive sparse attention mechanism by generalizing a sparsity-inducing activation function. Then a sparse communication graph in MARL is learned by graph neural networks based on this new attention mechanism. Through this sparsity structure, the agents can communicate in an effective as well as efficient way via only selectively attending to agents that matter the most and thus the scale of the MARL problem is reduced with little optimality compromised. Comparative results show that our algorithm can learn an interpretable sparse structure and outperforms previous works by a significant margin on applications involving a large-scale multiagent system.
\end{abstract}









\section{INTRODUCTION}
Reinforcement Learning (RL) has achieved enormous successes in robotics~\cite{kober2013reinforcement} and gaming~\cite{mnih2015human} in both single and multiagent settings. For example, deep reinforcement learning (DRL) achieved super-human performance in the two-player game Go, which has a very high-dimensional state-action space~\cite{silver2016mastering, silver2017mastering}. However, in multiagent scenarios, the sizes of the state space, joint action space, and joint observation space grow exponentially with the number of agents. As a result of this high dimensionality, existing multiagent reinforcement learning (MARL) algorithms require significant computational resources to learn an optimal policy, which impedes the application of MARL to systems such as swarm robotics~\cite{huttenrauch2017guided}. Thus, improving the scalability of MARL is a necessary step towards building large-scale multiagent learning systems for real-world applications. 

In MARL, the increase of complexity of finding an optimal joint policy, with respect to the number of agents, is a result of coupled interactions between agents \cite{decpomdp}. However, in many multiagent scenarios, the interactions between agents are quite sparse. For example, in a soccer game, an agent typically only needs to pay attention to other nearby agents when dribbling because agents far away are not able to intercept. The existence of such sparsity structures of the state transition dynamics (or the state-action-reward relationships) suggests that an agent may only need to attend to information from a small subset of the agents for near-optimal decision-making. Note that the other players that require attention might not be nearby, such as the receiver of a long pass in soccer. In such cases, the agent only needs to selectively attend to agents that ``matter the most". As a result, the agent can spatially and temporally reduce the scale of the planning problem. 

In large-scale MARL, sample complexity is a bottleneck of scalability~\cite{bu2008comprehensive}. To reduce the sample complexity, another feature we can exploit is the interchangeability of homogeneous agents: switching two agents' state/action will not make any difference to the environment. This interchangeability implies permutation-invariance of the multiagent state-action value function (a.k.a. the centralized $Q$-function) as well as interchangeability of agent policies. However, many MARL algorithms such as MADDPG~\cite{lowe2017multi}, VDN~\cite{sunehag2018value}, QMIX~\cite{rashid2018qmix} do not exploit this symmetry and thus have to learn this interchangeability from experience, which increases the sample complexity unnecessarily.

Graph neural network (GNN) is a specific neural network architecture in which permutation-invariance features can be embedded via graph pooling operations, so this approach has been applied in MARL~\cite{agarwal2019learning,khan2019graph, jiang2018graph} to exploit the interchangeability. As MARL is a non-structural scenario where the links/connections between the nodes/agents are ambiguous to decide, a graph has to be created in advance to apply GNN for MARL. Refs.~\cite{agarwal2019learning,khan2019graph, jiang2018graph}, apply ad-hoc methods, such as $k$-nearest neighbors, hard threshold, and random dropout to obtain a graph structure. However, these methods require handcrafted metrics to measure the closeness between agents, which are scenario-specific and thus not general/principled. Inappropriately selecting neighbors based on a poorly designed closeness metric could lead to the failure of learning a useful policy. 

While attention mechanisms \cite{vaswani2017attention} could be applied to learn the strength of the connections
between a pair of agents (i.e., closeness metric) in a general and principled way, such strengths are often dense, leading
to a nearly-complete computation graph that does not benefit scalability. The dense attention mechanism results from that the
softmax activation function operated on the raw attention logits generates a probability
distribution with full support. One solution to enforce a sparse graph is top $k$ thresholding~\cite{chorowski2015attention}, which keeps the $k$-largest attention scores and
truncates the rest to zero. However, this truncation is a non-differentiable operation that may cause problems for gradient-based optimization algorithms, such as those used in end-to-end training. Therefore, a sparse attention
mechanism that preserves the gradient flow necessary for gradient-based training is required. 

To address the non-differentiability issue in sparse attention mechanisms, we generalize sparsemax~\cite{martins2016softmax} and obtain a sparsity mechanism whose pattern is adaptive to the environment states. This sparsity mechanism can reduce the complexity of both the forward pass and the back-propagation of the policy and value networks, as well as preserving the end-to-end trainability in contrast to hard thresholding. With the introduction of GNN and generalized sparsemax, which can preserve permutation invariance and promote sparsity respectively, the scalability of MARL is improved.
 
The discussion so far was restricted to homogeneous agents and thus permutation-invariance is desirable. However, in heterogeneous multiagent systems or competitive environments, permutation invariance and interchangeability are no longer valid. For example, in soccer, switching positions of two players from different sides can make a difference to the game. To address this heterogeneity, GNN-based MARL must distinguish the different semantic meanings of the connections between different agent pairs (e.g. friend/friend relationship versus friend/foe relationship). We address this requirement by multi-relational graph convolution network~\cite{schlichtkrull2018modeling} to pass messages using different graph convolution layers on graph edge connections with different semantic meanings.

To summarize, we propose to learn an adaptive sparse communication graph within the GNN-based framework to improve the scalability of MARL, which applies to both homogeneous and heterogeneous multiagent systems in mixed cooperative-competitive scenarios.

\subsection{Related Work}

One of the existing works exploiting the structure in MARL is the mean-field reinforcement learning (MFRL)~\cite{yang2018mean} algorithm, which takes as input the observation and the mean action of neighboring agents to make the decision, and neglects the actions of all the other agents. This simplification leads to good scalability. However, the mean action cannot distinguish the difference among neighboring agents and the locality approximations fail to capture information from a far but important agent for optimal decision-making, which leads to sub-optimal policies. Multi-Actor-Attention-Critic (MAAC) is proposed in \cite{iqbal2018actor} to aggregate information using attention mechanism from all the other agents. Similarly,  \cite{agarwal2019learning, jiang2018graph, das2018tarmac} also employ the attention mechanism to learn a representation for the action-value function. However, the communication graphs used there are either dense or ad-hoc ($k$ nearest neighbors), which makes the learning difficult.

Sparse attention mechanisms were first studied by the natural language processing community in~\cite{martins2016softmax}, where sparsemax was proposed as a sparse alternative to the activation function softmax. The basic idea is to project the attention logits onto the probability simplex, which can generate zero entries once the projection hits the boundary of the simplex. While generalized sparse attention mechanisms were further studied in~\cite{niculae2017regularized, blondel2018learning,laha2018controllable}, they are not adaptive to the state in the context of MARL, in terms of the sparsity pattern.

Given this state of the art, the contributions of this paper are twofold. First, we propose a new adaptive sparse attention mechanism in MARL to learn a sparse communication graph, which improves the scalability of MARL by lowering the sample complexity. Second, we extend our GNN-based MARL to heterogeneous systems in mixed cooperative-competitive settings using multi-relational GNN. The evaluations show that our algorithm  significantly outperforms previous approaches on applications involving a large number of agents. This technique can be applied to empower large-scale autonomous systems such as swarm robotics.

\section{PRELIMINARIES}


\subsection{Multiagent Reinforcement Learning}
As a multiagent extension of Markov decision processes (MDPs), a Markov game is defined as a tuple $\langle N,S,\{O_i\}_{i\in N}, \{A_i\}_{i\in N}, \{r_i\}_{i\in N}, \gamma \rangle$, where $N = [1,\ldots,n]$ is a set of agent indices, $S$ is the set of state, $\{O_i\}_{i\in N}$ and $\{A_i\}_{i\in N}$ are the joint observation and joint action sets, respectively. The $i$th agent chooses actions via a stochastic policy $\pi_{\theta_i}: O_i \times A_i \to [0, 1]$, which leads to the next state according to the state transition function $\mathcal{T} : {S} × A_1 × \ldots × A_n \to S$. The $i$th agent also obtains a reward as a function of the state and agent’s action $r_i: S × \{A_i\}_{i\in N} \to \reals$, and receives a private observation correlated with the state $o_i: S\times \{A_i\}_{i\in N} \to O_i$. The initial states are determined by a distribution $\rho : S\to [0, 1]$. The $i$th agent aims to maximize its own total expected return $R_i = \sum_{t=1}^T\gamma^tr_i^t$, with discount factor $\gamma$ and time horizon $T$.

\subsection{Multi-head attention}\label{sec:attention}
The scaled dot-product attention mechanism was first proposed in \cite{vaswani2017attention} for natural language processing. An attention function maps the query and a set of key-value pairs to the output, which is the weighted sum of the values. The weight assigned to the each value calculated via a compatibility function of the query and the corresponding key. In the context of MARL, let $h_i,i\in N$ be the representation of the agents. Key, query and value of agent $i$ is defined as $K_i^{l} = W_K h_i^{l}\in \mathbb{R}^{d_K}$, $Q_i^{l} = W_Q h_i^{l}$ and $V_i^{l} = W_V h_i^{l}$, respectively with $W_K, W_Q$ and $W_V$ are parameter matrices. The output for agent $i$ is then
\begin{equation}\label{eq:hi}%
\text{Att}_i(h) = \sum_{j}w_{ij}V_j,
\end{equation}
where $w_{i\bullet} \in\mathbb{R}^{n}$, the $i$-th row of the weight matrix $w$, is defined as
\begin{equation}\label{eq:w}
w_{i\bullet} = \sigma_a\Big(\frac{(K_i)^TQ}{\sqrt{d_K}}\Big)
\end{equation}
with $\sigma_a$ being the softmax function in previous works of GNN-based MARL. The weight $w_{i\bullet}$ is dense as $\softmax_i(z) \ne 0$ for any vector $z$ and $i$.

To increase the expressiveness, multi-head attention is applied here via simply concatenating the outputs from a single attention function~\cite{vaswani2017attention}.



\subsection{Relational GNN}
In heterogeneous multiagent systems, different agent pair can have different relations, such as friend or foe in a two-party zero-sum game. As a result, information aggregation from agents with different relations should have different parameters. Work in \cite{schlichtkrull2018modeling} proposed relational graph convolutional network to model multi-relational data. The forward-pass update of agent $i$ in a multi-relational graph is as follows
\begin{equation}\label{eq:mr-gnn}%
h_i^{(l+1)} = \sigma\Big(\sum_{r\in\mathcal{R}}\sum_{j\in\mN_i^r} \frac{1}{c_{i,r}}W_r^{(l)} h_j^{(l)} + W_0^{(l)} h_i^{(l)}\Big),
\end{equation}
where $\mN_i^r$ denotes the set of neighbor indices of agent $i$ under relation $r\in\mathcal{R}$ and $c_{i,r}$ is a normalization constant. To distinguish the heterogeneity in MARL, similar to this convolution-based multi-relational GNN, we apply different attention heads on agent pairs with different relations.

\section{APPROACH}\label{sec:Approaches}
In this section, we present our approach to exploit the sparsity in MARL by generalizing the dense soft-max attention to adaptive sparse attention. Moreover, our approach to apply multi-relational attention mechanism for heterogeneous games involving competitive agents is also introduced. 
\subsection{Learning a communication graph via adaptive sparse attention}\label{sec:sparse}
The scaled dot-product attention is applied to learn the communication graph in MARL. If an attention weight between a pair of agents is zero, then there is no communication/message passing between them. Thus, the normalization function $\sigma_a(\bullet)$ in \eqref{eq:w} is critical to learn a communication graph. As usually used in the attention mechanism~\cite{vaswani2017attention} or classifications, $\sigma_a(\bullet)$ is usually set to be softmax, which cannot induce sparsity. We propose an adaptive sparse activation function as an alternative to softmax.

Let $x\in\mathbb{R}^{d}$ be the raw attention logits and $y$ be normalized attention strength in the ($d-1$)-dimensional probability simplex defined as $\Delta^d:=\{y\in\mathbb{R}^{d}|y\ge 0,\ones^T y=1\}$. We are interested in the mapping from $x\in\mathbb{R}^{d}$ to $y\in\Delta^d$. In other words, such a mapping can transform real weights to a probability distribution, i.e., the normalized attention strength between a pair of agents. The classical softmax, used in most attention mechanisms,  is defined component-wisely as 
\begin{equation}\label{eq:hi}
y_i = \softmax_i(x)=\frac{e^{x_i}}{\sum_{i=1}^{d}e^{x_i}}.
\end{equation}
A limitation of the softmax transformation is that the resulting probability distribution always has full support, which makes the communication graph dense, resulting in high complexity. In order to reduce the complexity, our idea is to replace the softmax activation function with a generalized activation function, which could adaptively be dense or sparse based on the state. To investigate alternative activation functions to softmax, consider the max  operator defined as 
\begin{equation}\label{eq:max}
\max(x):=\max_{i\in[d]}(x_i)=\sup_{y\in\Delta^d}y^Tx,		
\end{equation}
where $[d]=\{1,\ldots,d\}$. The second equality comes from that the supremum of the linear form over a simplex is always achieved at a vertex, i.e., one of the standard basis vector $\{e_i\}_{i\in[d]}$. As a result, the max operator puts all the probability mass onto a single element, or in other words, only one entry of $y$ is nonzero corresponding to the largest entry of $x$. For example, with $x=[0,t]\in \reals^2$, the probability distribution w.r.t. the logit $t$, i.e.,  $(\argsup_{y\in\Delta^d}y^Tx)_2$, is a step function,  as $(\argsup_{y\in\Delta^d}y^Tx)_2$ equals 1 if $t>0$ and $0$ otherwise. This discontinuity at $t=0$  of the step function is not amenable to gradient-based optimization algorithms for training deep neural networks.  
One solution to the discontinuity issue encountered in \eqref{eq:max} is to add a regularized $\Omega(y)$ in the max operator as 
\begin{equation}\label{eq:max}
\Pi_{\Omega}(x)=\argmax_{y\in\Delta^d}y^Tx + \gamma\Omega(y)		
\end{equation}
Different regularizers $\Omega(y)$ produce different mappings with distinct properties (see summary in Table~\ref{table:regularizers}). Note that with $\Omega(y)$ as the Shannon entropy, $\Pi_{\Omega}(x)$ recovers softmax. With the states/observations evolving, the ideal profile of $\Pi_{\Omega}(x)$ should be able to adapt the sparsity extent (controlled via $\gamma$) and the pattern (controlled via the selection of $\Omega(y)$) accordingly.

\begin{table}[t]
\caption{List of different regularizers and their corresponding mappings $y=\Pi_{\Omega}(x)$, where $x$ is the raw attention logits and $y$ is the probability distribution in  $\Delta^d$. }
\label{table:regularizers}
\begin{center}
\begin{tabular}{c||c|c|c} \hline
Entropy     & $\Omega(y)$            & $\Pi_{\Omega}(x)$    &Ref.\\ \hline \hline
\rule{0pt}{10pt}
Shannon     & $\sum_i y_i\log(y_i)$    & $\softmax_i(x)=\frac{e^{x_i}}{\sum_{i=1}^{d}e^{x_i}}$   &\cite{blondel2018learning}\\\hline
\rule{0pt}{10pt}
$l_2$ norm  & $-\frac{1}{2}\sum_i y_i^2$        & $\arg\min_{y\in\Delta^d}\|y-x\|^2$                            &\cite{martins2016softmax}\\\hline
\rule{0pt}{18pt}
 Tsallis     &      $ \hspace{-0.2cm}
                        \left\{
                            \begin{array}{ll}
                              \frac{\sum_i(y_i - y_i^{\alpha})}{\alpha(\alpha-1)}, &\hspace{-0.3cm} \alpha \ne 1\\
                              \sum_i y_i\log(y_i), & \hspace{-0.3cm} \alpha = 1
                            \end{array}
                          \right.
              $
  
                                    & No closed-form                                                 &\cite{correia2019adaptively}\\\hline
\rule{0pt}{16pt}
Generalized & $\displaystyle \frac{1}{q}\sum_i(y_i - \frac{\displaystyle e^{qy_i}-1}{\displaystyle e^q-1})$ 
                                    & No closed-form                                                 &\cite{kowalski2013concepts}\\
\hline
\end{tabular}
\end{center}
\end{table}



\begin{table*}[h]
\caption{List of different $G(x)$ and their resulting mappings  $\Pi_\Omega(x)$}
\label{table:Gx}
\begin{center}
\begin{tabular}{c||c|c|c}
\hline\\[-1em]
$\gamma G_i(x)$ & $\frac{e^{x_i}}{\sum_i e^{x_i}}$ & $\frac{{x_i^2}}{\sum_i {x_i^2}}$ & $x_i$\\\hline\hline
 $\Pi_\Omega(x)$&  softmax& softmax&sparsemax\\
\hline
Property &\makecell{Translation invariance \\  $\Pi_\Omega(x)= \Pi_\Omega(x+c\ones)$}   & \makecell{Scaling  invariance \\  $\Pi_\Omega(x)= \Pi_\Omega(cx)$}&\makecell{Translation invariance \\  $\Pi_\Omega(x)= \Pi_\Omega(x+c\ones)$ }\\\hline
Example &$\Pi_\Omega([100, 101])= \Pi_\Omega([0,1])$& $\Pi_\Omega([1,2])=  \Pi_\Omega([1, 2]\times 10^{-3})$&$\Pi_\Omega([100, 101])= \Pi_\Omega([0,1])$\\\hline
\end{tabular}
\end{center}
\end{table*}

Note that the Tsallis entropy and the generalized entropy in Table \ref{table:regularizers} do not have closed-form solutions~\cite{blondel2018learning}, which will increase the computational burden since iterative numerical algorithms will have to be employed. Sparsemax has a closed-form solution and can induce sparsity, but sparsemax is not adaptive and lacks flexibility as it is unable to switch from one sparsity pattern to another when necessary. We aim to combine the advantages and avoid the disadvantages using this new formulation 
\begin{equation}\label{eq:my_sparsemax}
\Pi_{\Omega}(x)=\argmin_{y\in\Delta^d}||y-\gamma G(x)||^2,				
\end{equation}
with $G(x):\reals^d\to\reals^d$ and $\gamma$ being a learnable neural network and a scalar, respectively. By choosing different $G(x)$, $\Pi_{\Omega}(x)$ can exhibit different sparsity patterns including softmax and sparsemax. With $G(x)$ fixed, the parameter $\gamma$ can control how sparse the output could be, similar to the temperature parameter in softmax. The summary in Table \ref{table:Gx} shows that \eqref{eq:my_sparsemax} will lead to a general mapping and can combine properties such as translation and scaling invariance adaptively. Work in~\cite{laha2018controllable} proposed sparse-hourglass that can adjust the trade-off between translation and scaling invariance via tunable parameters. However, it is unclear under which circumstances one property is more desirable than the other, so there is little to no prior knowledge on how to tune such parameters. In contrast, our formulation in \eqref{eq:my_sparsemax} can balance such trade-off via learning $G(x)$ and $\gamma$ while work in~\cite{laha2018controllable} is based on a fixed form of $G(x)$ with tunable parameters.


While we can let the neural network learn $G(x): \reals^d \to \reals^d$ without any restrictions, there is indeed prior knowledge that we can apply, e.g., monotonicity. It is desired to keep the monotonicity of  $\Pi_{\Omega}(x)$, i.e., $\forall x_i>x_j, (\Pi_{\Omega}(x))_i>(\Pi_{\Omega}(x)_j$, as larger attention logit should be mapped into larger attention strength. As sparsemax is monotonic, this requires that $\forall x_i>x_j, G_i(x) > G_j(x)$, or in other words, the order of the input of $G(x)$ coincides with that of the output. To keep this property, $G(x)$ is designed component-wisely as $G_i(x)=\psi(\phi_1(x_i),\sum_i \phi_2(x_i))$, with $\psi: \reals^2 \to \reals^1,  \phi_1,\phi_2:\reals^{1}\to \reals^1$ are neural networks with hidden layers. Note that $G_i(x)$ should be coupled with all of the entries of $x$ instead of be a univariate function only depending on $x_i$, as demonstrated in Table II. As the second argument of $\psi$ (i.e., $\sum_i \phi_2(x_i)$) is invariant to $G_i(x), \forall i \in [d]$, the order preserving of $G(x): \reals^d \to \reals^d$ is equivalent to the monotonicity of $\psi(\bullet)$ and $\phi_1(\bullet)$. In order to keep this monotonicity, we enforce all the weights of the networks $\psi$ and $\phi_1$ to be positive~\cite{dugas2009incorporating}, by applying an absolute value function on the weights. This architecture can accelerate the learning process with extra prior knowledge, as it is monotonic by design.

\subsection{Message passing in MARL via GNN}
We will present 
how the information is aggregated to learn a representation for per-agent value/policy network using a graph neural network. The scaled dot-product attention mechanism (Section~\ref{sec:attention}) with our generalized sparsemax as the activation function, denoted as \textit{sparse-Att}, is applied to learn a communication graph and pass messages through the connections in the graph.

We start with homogeneous multiagent system, where the relation between any agent pair is identical. A graph is defined as $\mathcal{G}:= (\mathcal{V}, \mathcal{E})$, where $v_i\in \mathcal{V}$ represent an agent and the cardinality of $\mV$ is $|\mV|$. Moreover, $e_{ij}\in \mathcal{E}$ is $1$ if agent $i$ and $j$ can communicate directly (or agent $j$ is observable to agent $i$), and $0$ otherwise. This is a restriction on the communication graph and $\mathcal{E}$ is the set of all possible edges. Then sparse-Att aims to learn a subset of $\mathcal{E}$ via induced sparsity without compromising much optimality. For agent $i$, let $U_i = f_a(X_i)$ and $E_i$ be its observation and entity encoding respectively, where $X_i, i\in \mathcal{V}$ is the local state and $f_a$ is a learnable agent encoder network. Then the initial observation embedding of agent $i$, denoted as $h_i^{(1)}$, is  
\begin{equation}\label{eq:hi}%
h_i^{(1)} = f_{mp}(U_i \| E_i),
\end{equation}
where $f_{mp}$ is another learnable network and the operator $\|$ denotes concatenation. Then at hop $l$ ($l$-th round of message passing), agent $i$ aggregates information from its possible neighbors belonging to the set $\mathcal{N}=\{j\in\mathcal{V}|e_{ij}=1\}$ as follows
\begin{equation}\label{eq:homo_mp}%
h_i^{(l+1)} = f_{mp}\Big(h_i^{(l)}\|\text{sparse-Att}_i^\mN(h^{(l)})\Big).
\end{equation}
With $l\ge2$, the multi-hop message passing can enable the agent to obtain information from beyond its immediate neighbors. In the message aggregation from all of the agents $\mN$, identical parameters are used in $\text{sparse-Att}_i^\mN$, which enforces the permutation-invariance. This property is desirable because homogeneous agents are interchangeable.

However, interchangeability is no longer applicable to heterogeneous systems or mixed cooperative-competitive environment. For example, with $\mV_1, \mV_2\subseteq \mV$ being a two-team partition of $\mV$, agents cooperate with other agents from the same team but compete against agents from the other team. For agent $i\in\mV_1$, its teammate neighborhood and enemy neighborhood are  $\mathcal{N_+}=\{j\in\mathcal{V}_1|e_{ij}=1\}$ and $\mathcal{N_{-}}=\{j\in\mathcal{V}_2|e_{ij}=1\}$, respectively. The edges connecting teammates and enemies are called positive and negative edges. Then based on multi-relational GNN, agent $i$ aggregates information at hop $l$ in the following way
\begin{equation}\label{eq:hetero_mp}%
h_i^{(l+1)} = f_{mp}\Big(h_i^{(l)} \|\text{sparse-Att}_i^{\mN_+}(h^{(l)}) \|\text{sparse-Att}_i^{\mN_-}(h^{(l)})\Big), \nonumber
\end{equation}
where $\text{sparse-Att}_i^{\mN_+}$ and $\text{sparse-Att}_i^{\mN_-}$ are different attention heads. Additionally, balance theory~\cite{heider1946attitudes} suggests that ``the teammate of my teammate is my teammate" and ``the enemy of my enemy is my teammate." In a two-team competitive game, any walk (a sequence of nodes and edges of a graph) between an agent pair in the communication graph, comprising of both positive and negative edges, will lead to the same relation between the agent pair~\cite{easley2012networks}. This property eliminates the ambiguity that the information aggregated from the same agent (but different walk) might have a different teammate/enemy property.

\begin{figure}[t]
  \centering
  \includegraphics[scale=0.6]{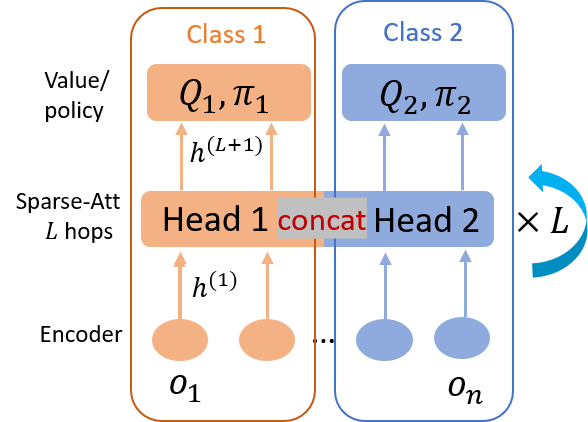}
  \caption{\small Our sparse-Att framework consists of three modules: encoder, multi-relational sparse attention mechanism, and value/policy network, with homogeneous agents sharing all parameters. Agents employ different attention heads to aggregate information alongside connections with different semantic meanings, followed by a concatenation. $L$ is the number of the message-passing rounds; see \eqref{eq:homo_mp}. ``concat" denotes the concatenation operation. Here only two classes (shown in red and blue) of heterogeneous agents are shown for simplicity. }
  \label{fig:illustration}
\end{figure}

The proposed algorithmic framework is illustrated in Fig.~\ref{fig:illustration}. After $L$ rounds of message passing, each agent has an updated encoding $h_i^{(L+1)}$. This encoding is then fed into the value network and the policy network, which estimate the state value and a probability distribution over all possible actions, respectively. As homogeneous agents are interchangeable, they share all of the parameters, including entity encoding, policy, value and message passing. Proximal policy gradient (PPO, \cite{schulman2017proximal}) is employed to train the model in an end-to-end manner. As only local information is required, the proposed approach is decentralized. Moreover, our approach maintains the transferability of GNN-based approaches as all the network dimensions are invariant to agent/entity number in the system.

\section{EXPERIMENTS}
\subsection{Task description}

The proposed algorithm is evaluated in three swarm robotics tasks: Coverage, Formation, and ParticleSoccer~\cite{csahin2004swarm}, first two of which are cooperative and the third is competitive. The tasks are simulated in the Multiagent Particle Environment\footnote{https://github.com/openai/multiagent-particle-envs}(MAPE~\cite{lowe2017multi}). The agents in MAPE can move in a 2-dimensional space following a double integrator dynamic model. The action space of the agents is discretized, with each agent can accelerate/decelerate in both $X$  and $Y$ direction. The three tasks are briefly introduced as follows.

\begin{figure}[t]
\centering
\begin{subfigure}{0.32\columnwidth}
\includegraphics[width=\linewidth, frame]{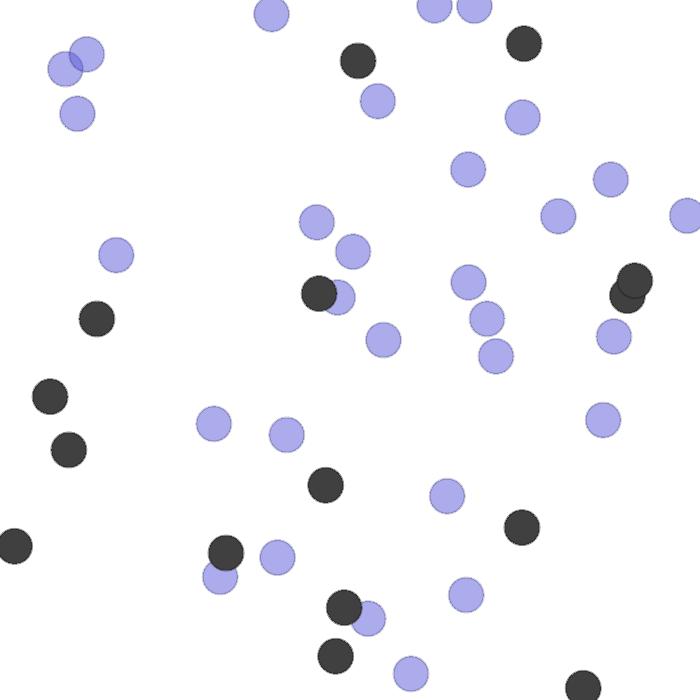}
\caption{Coverage}
\label{fig:spread}
\end{subfigure}
\begin{subfigure}{0.32\columnwidth}
\includegraphics[width=\textwidth, frame]{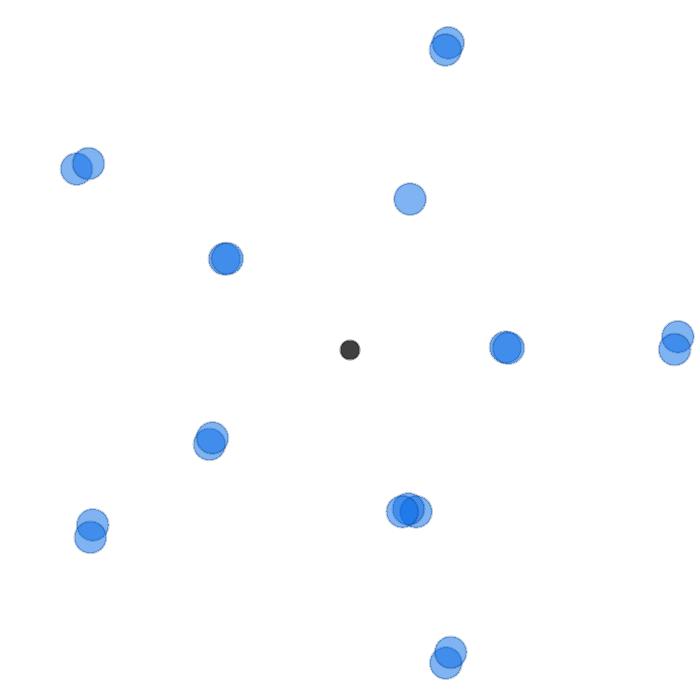}
\caption{Formation}
\label{fig:formation}
\end{subfigure}
\begin{subfigure}{0.32\columnwidth}
\includegraphics[width=\textwidth, frame]{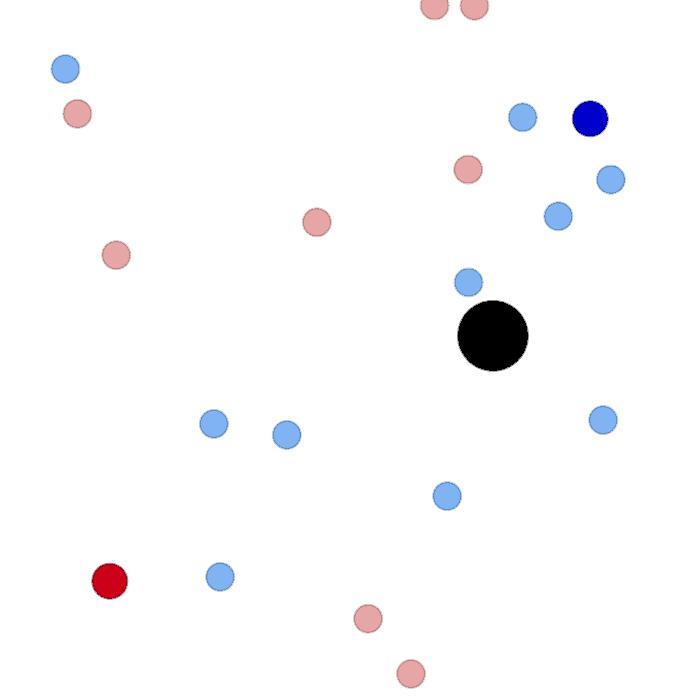}
\caption{ParticleSoccer}
\label{fig:soccer}
\end{subfigure}
\caption{Three different simulation tasks used in this work.}%
\label{Fig:tasks}%
\end{figure}

\textbf{Coverage:} There are $n_A$ agents (light purple) and $n_L$ landmarks (black) in the environment (see illustration in Fig. \ref{fig:spread}). The objective for the agents is to cover the landmarks with the smallest possible number of timesteps. Agents are not assigned to reach a certain landmark, but instead, have to figure out the assignment via communication such that the task can be finished optimally.

\textbf{Formation:} There are $n_A$ agents (blue) and $1$ landmarks (black) in the environment (see illustration in Fig. \ref{fig:formation}), with $n_A$ being an even natural number. The agents need to split into two sub-teams of equal size, with each of them building a formation of a regular pentagon. The two regular pentagons with different sizes are both centered at the landmark. 

\textbf{ParticleSoccer:} There are $n_A$ agents and 3 landmarks in the environment (see illustration in Fig. \ref{fig:soccer}), with the bigger landmark as a movable ball and the two smaller ones as a fixed landmark. A team wins the game via pushing the black ball to the opponent team's goal. The goal color of the light blue (red, resp.) team is blue (red, resp.). 

\subsection{Implementation specifications}
The agent encoder $f_a(\bullet)$ and the entity encoder take input the $4$-dimensional agent states and $2$-dimensional entity states, respectively. The queries, keys, and values in all of the sparse attention mechanism are $128$-dimensional. The communication hop is $L=2$. All neural networks are fully connected with the ReLU activation function. In the sparsity-promoting function \eqref{eq:my_sparsemax}, $\phi_1, \phi_2$  and $\psi$ all have one hidden layer with dimensions being $16$, $16$ and $64$, respectively. The absolute value function is used to keep the weights of the monotonicity-preserving neural network positive.

Evaluation is performed every $320$ episodes and PPO update is executed for $4$ epochs after collecting experience of $4096$ timesteps.

\subsection{Results}
In the cooperative scenarios i.e., Coverage and Formation, two metrics are used to evaluate the algorithms. The first is the average reward per step and the second is the task success rate. Higher means better performance for both metrics.

We compare our algorithms with two baselines: GNN-based MARL with dense attention mechanism~\cite{agarwal2019learning} and MAAC~\cite{iqbal2018actor}. These two algorithms are considered to be strong baselines as they reported advantageous results against algorithms including MADDPG~\cite{lowe2017multi}, COMA~\cite{foerster2018counterfactual}, VDN~\cite{sunehag2017value} and QMIX~\cite{rashid2018qmix}. Public repositories\footnote{https://github.com/sumitsk/matrl.git}\footnote{ https://github.com/shariqiqbal2810/MAAC} are used for comparison. As both repositories also apply their algorithms on MAPE, the default hyperparameters are used for comparison.

\begin{figure}[t]
  \centering
  \includegraphics[width=0.9\columnwidth]{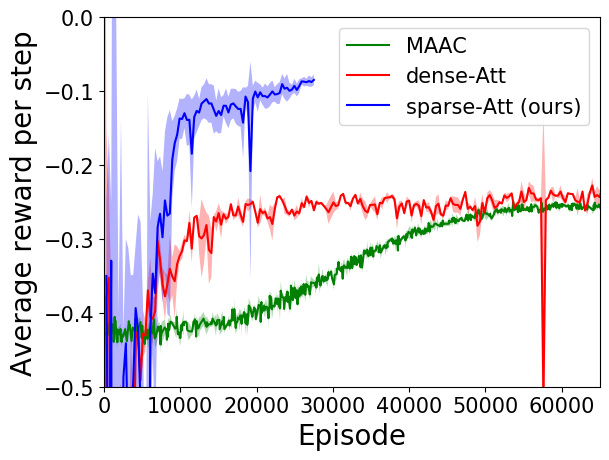} \vspace*{-.1in}
  \caption{Reward comparison of our algorithm against two baselines for the Coverage task.}
  \label{fig:spread_reward}
 %
 \vspace*{-.01in}
  %
  \centering
  \includegraphics[width=0.9\columnwidth]{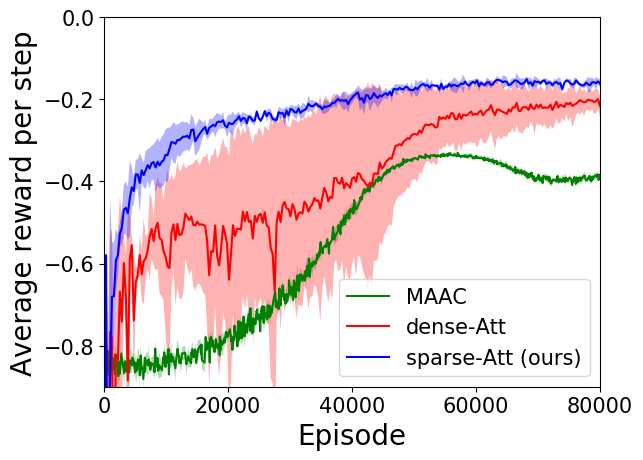}\vspace*{-.1in}
  \caption{Reward comparison of our algorithm against two baselines for the Formation task.}
  \label{fig:formation_reward}
\end{figure}

\begin{figure}[t]
\centering
\begin{subfigure}{0.485\columnwidth}
\centering
\hspace{-5cm}
\includegraphics[width=\textwidth]{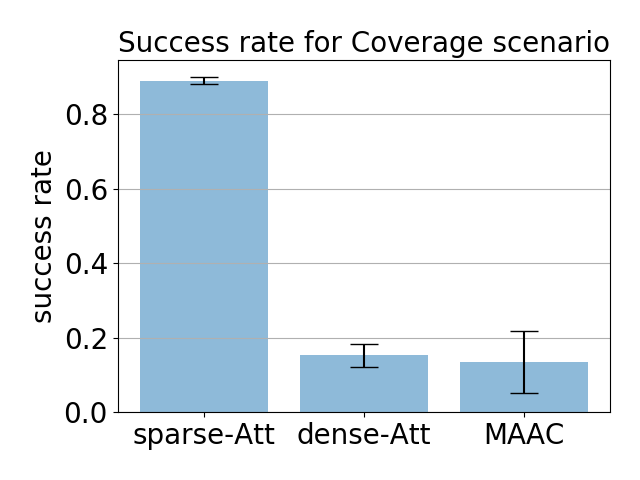}
\hspace{-5cm}
\caption{Coverage}
\label{fig:coverage_comparison}
\end{subfigure}
\begin{subfigure}{0.485\columnwidth}
\centering
\hspace{-5cm}
\includegraphics[width=\textwidth]{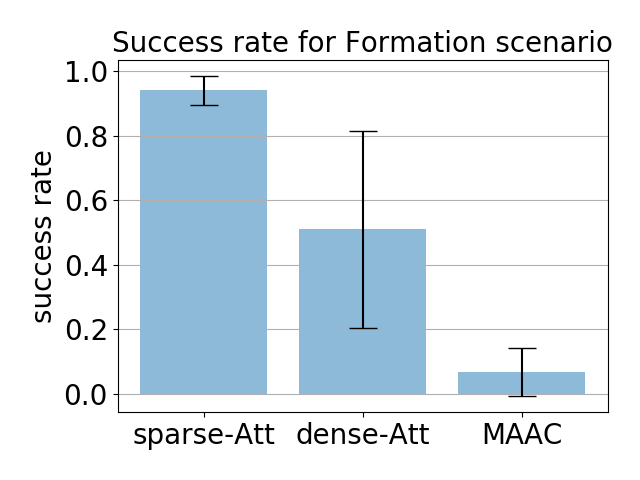}
\hspace{-5cm}
\caption{Formation}
\label{fig:formation_comparison}
\end{subfigure}
\caption{Performance comparison of three algorithm on two scenarios. Multiple policies learned from each algorithm are evaluated and the mean/standard deviation are plotted.}
\label{fig:tasks_comparison}
\end{figure}

In simulation, we set $n_A=30$ and $n_A=20$ for Coverage and Formation, respectively. Fig. \ref{fig:spread_reward} and Fig. \ref{fig:formation_reward} demonstrated that our algorithm can achieve higher rewards than the two baselines with fewer episodes. This validates that sparse-Att can accelerate the learning process via aggregating information from agents that matter the most. Moreover, in terms of the second metric, i.e., success rate, our algorithm consistently outperforms the two baselines by a significant margin (with a much smaller variance), as shown in Fig.~\ref{fig:tasks_comparison}. The evaluations of both metrics for two scenarios provide strong support for the advantages of our algorithm.

For the competitive ParticleSoccer task, we set $n_A=20$ with both red team and blue team of size $\frac{n_A}{2}=10$. As this task is competitive, the above two metrics are no longer applicable. Instead, we let the red (blue, resp.) play against a blue (red, resp.) team from another algorithm. Table \ref{table:SoccerComparison} presents the results of the inter-algorithm competition. The overall score of each algorithm equals the sum of the winning evaluation episodes of its red team and blue team playing against blue and red team respectively from other algorithms. The overall scores in Table \ref{table:SoccerComparison} show that our algorithm can learn strong policies.

\begin{table}[t]
\caption{Evaluation of three algorithms in the competitive ParticleSoccer task. Each pair is evaluated for $50$ episodes and the $(\bullet,\bullet,\bullet)$ in each cell denotes the number  of red team winning episodes, blue team wining episodes and the draw episodes. A draw means that neither team scores within a given episode length. $\text{win}_{\text{red}}$ and $\text{win}_{\text{blue}}$ are the winning episodes of the red and blue team, respectively when competing against blue and red team from other algorithms.}
\label{table:SoccerComparison}
\begin{center}
\begin{tabular}{c||c|c|c||c}
\hline\\[-1em]
\backslashbox{Red}{Blue}              & \makecell{sparse-Att \\  \textbf{(ours)}} & dense-Att & MAAC    & $\text{win}_{\text{red}}$\\
\\[-1em]\hline\hline
\makecell{sparse-Att \\  \textbf{(ours)}}       & \cellcolor{blue!25}$(48,0,2)$    &$(15,0,35)$     & $(26,0,24)$ & $41$\\
\hline
 dense-Att       &  $(9,1,40)$   & \cellcolor{blue!25}$(5,0,45)$   & $(3,0,47)$  & $11$\\
 \hline
 MAAC           &  (7,0,43)     &  (2,0,48)                 & \cellcolor{blue!25}$(3,0,47)$  & $9$         \\
\hline
$\text{win}_{\text{blue}}$ &   $-15$   &$-17$  &$-29$ & N/A\\
\hline
\hline
& \makecell{sparse-Att \\  \textbf{(ours)}} & dense-Att & MAAC    & \\
\hline
\makecell{overall scores:\\  $\text{win}_{\text{red}} + \text{win}_{\text{blue}}$} &$\mathbf{26}$ &$-6$ &$-20$ &\\
\hline
\end{tabular}
\end{center}
\end{table}

\subsection{Interpretability of the sparse communication graph}

Let us proceed by considering the inherent sparity in Formation and ParticleSoccer. 
As mentioned in the description of the Formation scenario, the formation of each pentagon is related to half of the agents, while the sub-team assignments need to be learned. In the implementation, the reward is set to require that the first $\frac{n_A}{2}$ agents closest to the landmark build the formations of the inner pentagon and the remaining $\frac{n_A}{2}$ agents to build the formations of the outer pentagon. With the convergence of the learning algorithm, once a sub-team partition is learned to complete the two sub-tasks, the learned agent indexing of each team should not vary due to the distance sorting and the two pentagons are relatively far away. As a result, the reward to complete each sub-task is only related to the corresponding sub-team and hence the two sub-teams are decoupled from each other. The adjacency matrix of the learned communication graph shown in Fig.~\ref{fig:sparse_comm_graph} validates that the inter-team communication is very sparse. This adjacency matrix is up to row/column permutation as indexing of each sub-team is learned without being known as a prior. Moreover, in a sub-team, the algorithm learns a communication graph similar to a star-graph. It can be understood that each sub-team selects a leader. As a star-graph is a connected graph with possibly minimum edges, this communication protocol is both effective and efficient. Also, the length of the path between any agent pair in a star graph is no greater than $2$, which echos the two-hop communication ($L=2$) we used in the simulation. That is because due to the two-hop message-passing, the agents can eventually communicate with agents as far as two edges away, which includes all of the agents in a star graph. Note that the sparsity on the diagonal entries of the communication graph does not mean that the agent's own information is neglected, as it is separately concatenated; see \eqref{eq:homo_mp}.

\begin{figure}[t]\label{fig:sparse_comm_graph}%
\centering
\begin{subfigure}{0.485\columnwidth}
\vspace{-3cm}
\centering
\hspace{-5cm}
\includegraphics[clip,trim=100 0 100 0, width=2in]{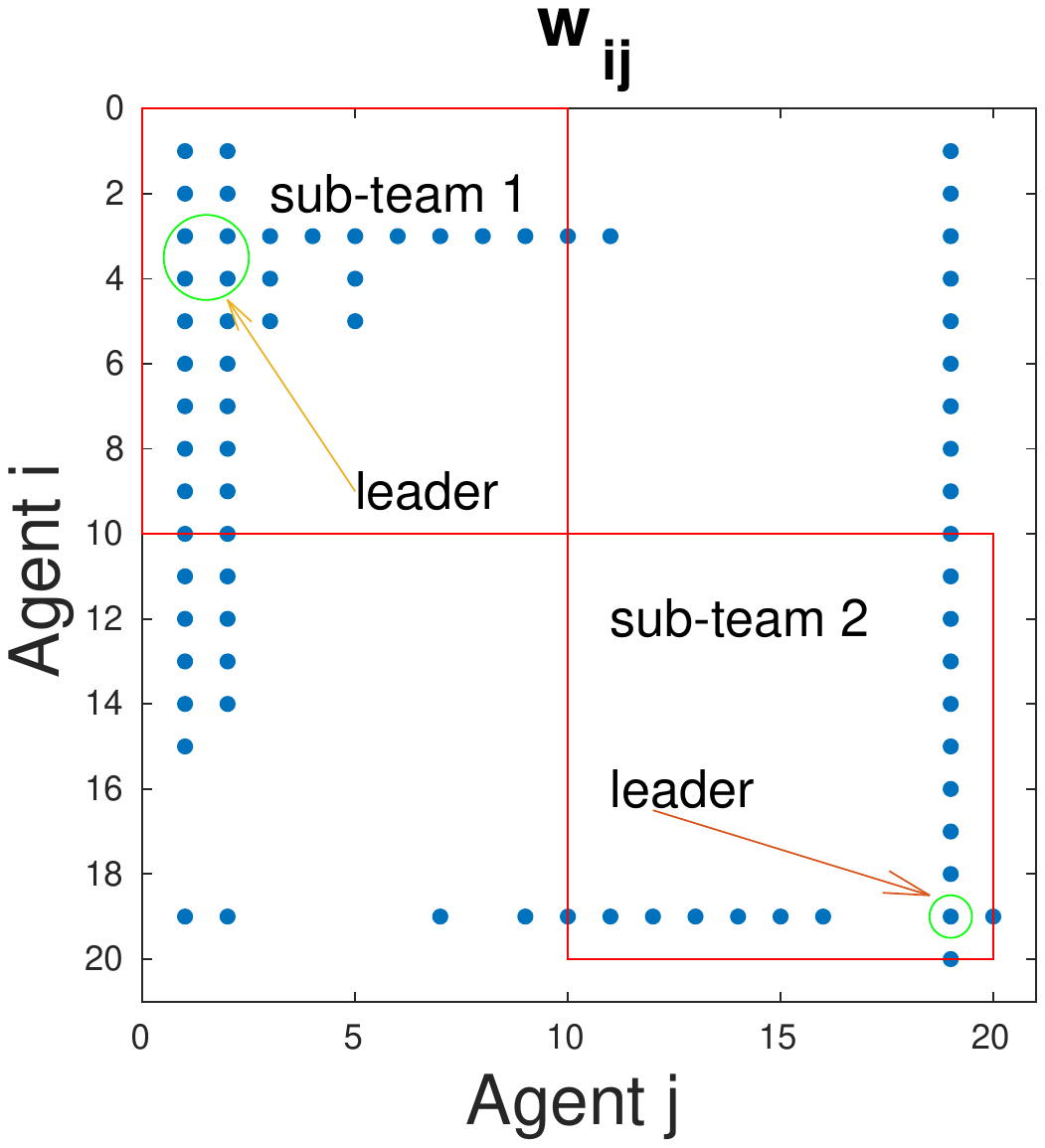}
\hspace{-5cm}
\vspace{-2.8cm}
\caption{Coverage}
\label{fig:sparse_comm_graph}
\end{subfigure}
\begin{subfigure}{0.485\columnwidth}
\vspace{-3cm}
\centering
\hspace{-5cm}
\includegraphics[clip,trim=100 0 100 0, width=2in]{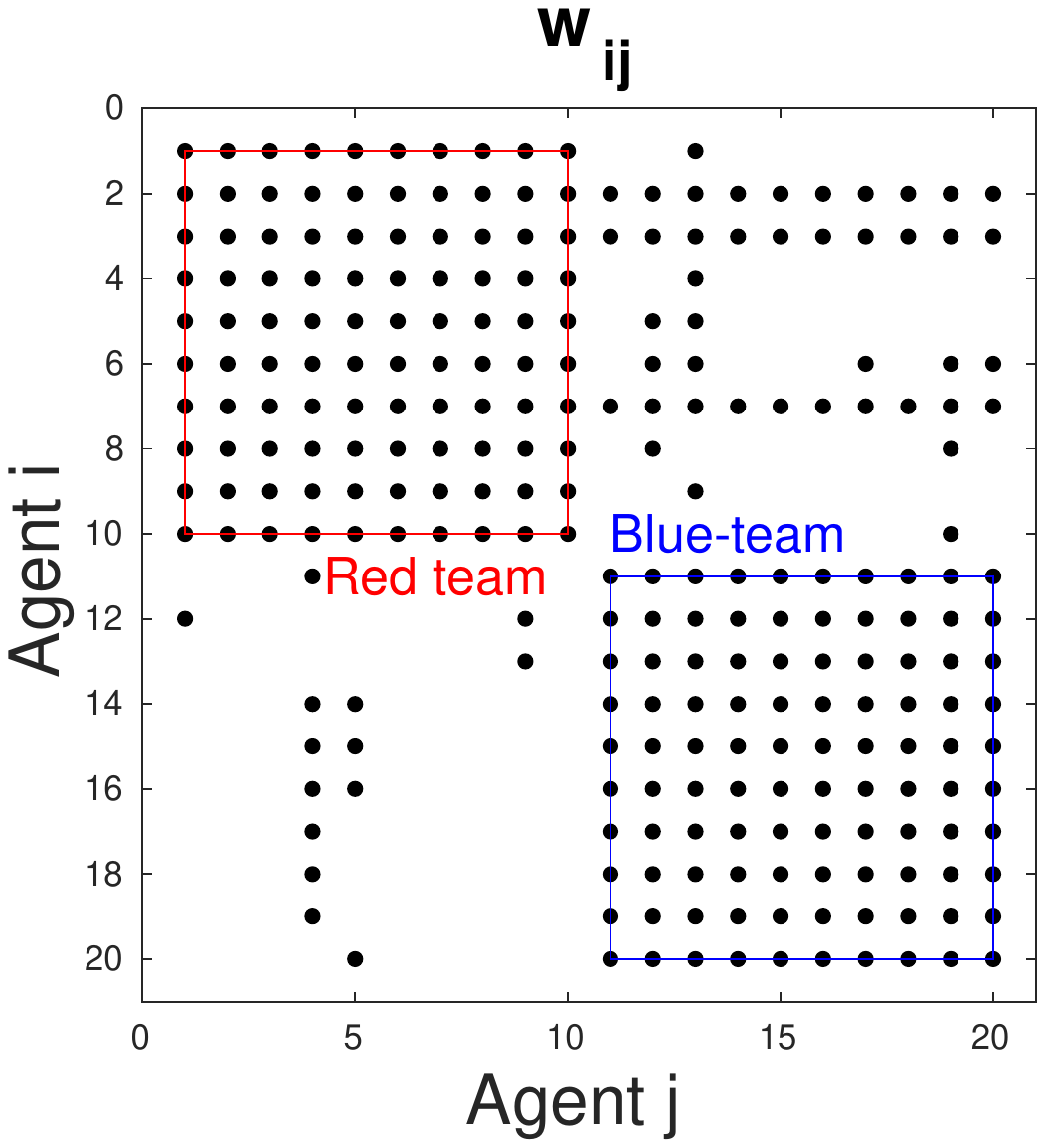}
\hspace{-5cm}
\vspace{-2.8cm}
\caption{Formation}
\label{fig:sparse_comm_graph_compet}
\end{subfigure}
\caption{Sparse communication graph for two scenarios. For the Coverage, our sparse-Att learns to split into two sub-team as desired and the learned sparse star-like communication graph makes communication both effective and efficient. In the ParticleSoccer, sparse-Att learn to pay more attention to teammates and a necessary subset of enemies.}
\end{figure}

Also, in the ParticleSoccer scenario, from each team's perspective, agents need to coordinate tightly within the team to greedily push the ball to the other team's goal while only attending to a small number of agents from the other team. This leads to dense intra-team communication but relatively sparse inter-team communication. This is validated by the approximately block-diagonal adjacency matrix of the learned communication graph in Fig. \ref{fig:sparse_comm_graph_compet}.



\section{CONCLUSIONS and FUTURE WORK}
This paper exploits sparsity to scale up Multi-Agent Reinforcement Learning (MARL), which is motivated by the fact that interactions are often sparse in multiagent systems. We propose a new general and adaptive sparsity-inducing activation function to empower an attention mechanism, which can learn a sparse communication graph among agents. The sparse communication graph can make the message-passing both effective and efficient such that the scalability of MARL is improved without compromising optimality. Our algorithm outperforms two baselines by a significant margin on three tasks. Moreover, for scenarios with inherent sparsity, it is shown that the sparsity of the learned communication graph is interpretable. 

Future work will focus on combining evolutionary population curriculum learning and graph neural network to further improve the scalability. In addition, robust learning against evolving/learned adversarial attacks is also of great interest.






\section*{ACKNOWLEDGMENTS}
Research is supported by Scientific Systems Company, Inc. under research agreement $\#$ SC-1661-04. 
Authors would like to thank Dong-Ki Kim, Samir Wadhwania and Michael Everett for their many useful discussions and Amazon Web Services for computation support.





\bibliographystyle{IEEEtran}
\bibliography{reference}
\balance

\balance

\end{document}